# Bayesian estimation of topological features of persistence diagrams

Asael Fabian Martínez

Universidad Autónoma Metropolitana, Unidad Iztapalapa, Mexico City, Mexico

`fabian@xanum.uam.mx`

**Abstract**

Persistent homology is a common technique in topological data analysis providing geometrical and topological information about the sample space. All this information, known as topological features, is summarized in persistence diagrams, and the main interest is in identifying the most persisting ones since they correspond to the Betti number values. Given the randomness inherent in the sampling process, and the complex structure of the space where persistence diagrams take values, estimation of Betti numbers is not straightforward. The approach followed in this work makes use of features' lifetimes and provides a full Bayesian clustering model, based on random partitions, in order to estimate Betti numbers. A simulation study is also presented.

## 1 Introduction

Geometrical and topological methods are modern tools for analyzing highly complex data [5, 8, 12]. While geometrical techniques capture quantitative information in data, topology reveals qualitative information. Both are useful to uncover patterns and relationships in data and, together with statistical and computational concepts and tools, sometimes complementary, form a powerful set of methods for analyzing modern data.

In particular, topological data analysis (TDA) is a modern field of applied mathematics with considerable interest and activity during the last two decades. It is a collection of tools in the field of data analysis that lies at the intersection of Algebraic Topology, Computational Geometry, Computer Science and Statistics. The main goal of TDA is to use ideas and results from Geometry and Topology to develop tools for revealing and describing relevant features of data objects with an intrinsic and complex structure. Given a cloud point data (dataset, in the terminology in this field), TDA methodologies are useful to understand their underlying space, so we can infer about its shape regardless of the choice of coordinates, deformations or presentations. This is done by estimating topological invariants related to the space, which capture the intrinsic clusters and connections among the clusters present in the cloud point data as well as other connectivity information, including the classification of loops and higher dimensional surfaces within the space.

Since TDA is in its core an issue in inference, to discover an unknown structure feature based on sampled cloud point data, some natural questions arise for statisticians. One is how it differs from classical cluster analysis. As pointed out by Carlsson, "TDA uses cluster analysis in building its networks, and builds on cluster analysis to provide additional precision in the taxonomies that are created" [9].

There are several TDA methodologies. One of the most common focuses on what is called *persistent homology*, which makes use of algebraic tools in order to discover topological features of data, where the so-called *Betti numbers* codify the number of $k$-dimensional holes present in the underlying space for different values of $k$. The method was introduced by [17] and its theory has been further considered in, for example, [8], [16], [26], [46] and [64]. Furthermore, persistent homology has been applied in a variety of fields including interconnectedness in the banking system [14], manufacturing systems [27] and computational biology with industrial and medical engineering applications [25, 59]. Other notable applications have been in data analysis [2, 8, 32, 43, 55, 60], image analysis [10, 22, 52], detection of subtypes of cancer [1, 42], analysis of brain artery trees [3], virus evolution [11, 30, 47], complex networks [29], language processing [63], sensor networks [15], spectroscopy [44], and soil science [50], among others.

Roughly speaking, persistent homology can be described as follows. Consider a finite cloud point data with pairwise distances between their points, select a scale $\epsilon > 0$ and join all points at a distance not more than $2\epsilon$. This gives an indication of a possible topological feature in the data, in particular an initial cluster classification; however, this depends on the scale, which could be hard to choose if the data are high-dimensional. Persistent homology examines data over all scales. The output of this computation is a summary called the *persistence diagram*, a set of pairs (*birth*, *death*). This and the *persistence barcode* encode life spans of topological features from their birth to death, from where one can choose, visually, pairs with a long persistence lifetime (*death–birth*). The procedure is performed over all the dimensions of the data. Topological features of short-scale duration are referred as *topological noise*, whereas the rest are called *topological signal*. Thus, discriminating between these features is of great interest since the topological signal is closely related with the Betti numbers of the underlying space.

A second natural question for statisticians is how randomness and uncertainty are handled. Any topological feature extracted from the cloud point data has random variation that needs to be taken into consideration for meaningful topological inference, and this is exactly the subject matter of Statistics. As referred in [45], practitioners of TDA often have backgrounds in pure topology and are not well-versed in statistical approaches to data analysis and there are several challenges for statistically interpreting results in applications of persistent homology, as few statistical tools are currently available. Conversely, TDA methodology has been slow to spread and develop within the statistical community and literature. In the setting of Statistics, persistence diagrams and barcodes are data summaries, or *statistics* based on cloud point data. Thus, notions of probability models for data and sampling distributions apply. A statistical approach to persistent topology was first considered in [7] where, among other aspects, theoretical persistence barcodes for parametric probability distributions on manifolds, as those considered in directional data, are derived, and then persistence barcodes are estimated using statistical inference principles, like maximum likelihood.

A general challenge for the statistical analysis of persistence diagrams is first to consider probability distributions on the set of persistence diagrams. This set is geometrically very complex, and it is difficult to consider parametric distributions for those diagrams, as to allow practical statistical inference. Moreover, persistence diagrams are not in an vector space and therefore one cannot use basic statistical tools like means, variance and moments, but rather Fréchet means [see 38, 40, 53]. As an alternative, *persistence landscapes* are proposed in [6]. They give topological summaries belonging to a space of functions and therefore law of large numbers and central limit theorems in function Banach spaces can be used to perform ad-hoc, non-parametric, classical statistical inference. Under a Bayesian framework, persistence diagrams are modeled in [35] through Poisson point processes for hypothesis testing in classification. On the other hand, for potential use in Statistics, there is the probabilistic limit theorem for persistence diagrams, which is an area of increasing interest in the framework of stochastic topology [4, 28, 31, 61, 62].

Another role for Statistics and Probability in persistent homology lies in the problem of disentangling topological noise from topological signals in persistence diagrams. In this direction, [19] and [13] proposed the construction of confidence sets for persistence diagrams, using geometric, statistic and probabilistic tools like kernel estimates, concentration inequalities, bootstrap, empirical processes, distance to measure and kernel distance. Those sets show the topological noise corresponding to all topological dimension homologies simultaneously. The confidence sets in [13] are robust, and it is also indicated how to construct confidence sets for particular dimension homologies, using the so-called bottleneck bootstrap.

The methodology presented in this work focuses on this second problem of identifying the topological signal. The interest of this topological feature lies in the fact that the quantity of signal features corresponds to the Betti number for each fixed dimension homology $k$. Hence, given an estimate of these numbers, it will be possible to understand the underlying space where the cloud point data live.

For every homology level $k$, the lifetimes computed from the persistence diagram are used to identify the topological noise and signal. Since, roughly speaking, the noise is of small magnitude when compared to the signal, and both of them inherit the randomness of the sampled cloud point data, the disentanglement of both features can be posed as a problem of outlier detection. The number of outlier lifetimes, thus, will correspond to the estimated Betti number. The detection of outliers is done by means of a full Bayesian model based on random partitions. The specific support for the random partition will allow to identify the topological signal from the groups containing the largest lifetime values.

The paper is organized as follows. Section 2 briefly explains persistent homology and how it is related to clustering and provides more information about the underlying sample space. Section 3 presents the proposed Bayesian model for estimating the Betti numbers via outlier detection. A simulation study is also performed in Section 4, and Section 5 contains some final remarks and future work.

## 2 Topological data analysis in a nutshell

TDA and Statistics, at first glance, tackle the same problem of clustering, as already mentioned. This makes it a good starting point to better understand where TDA, in particular persistent homology, differs; we can also have a clear picture of its potential. For a more comprehensive treatment of TDA foundations, we refer the reader to [41], [21], and [57], among others. Also, the work [45] overviews and compares the various methods available for computing persistent homology.

Cluster analysis aims to gather a set of items according to some similarity conditions. This set is partitioned into non-empty subsets, called groups or clusters, in such a way that all items in the same group are more similar among them than those in any other group. The degree of similarity is usually quantified according to some probability model or a specific distance function. Under the distance-based approach, a common method is the agglomerative hierarchical clustering with single linkage. Suppose we use the Euclidean distance, $d$. Given a cloud point data, we can build a graph using these points as vertices and its edges defined as follows: for a fixed $\epsilon \geq 0$, an edge $xy$ is added if and only if $d(x, y) \leq 2\epsilon$ for any two different vertices $x$, and $y$. Clusters are obtained from the connected components of the resulting graph. Figure 1a illustrate this procedure. The challenging part is to choose an appropriate cut-off value for $\epsilon$, since when $\epsilon = 0$, there will be as many clusters as observations, and when $\epsilon$ is large enough, there will be a single cluster containing all the observations.

Regarding the cut-off value, the persistent homology approach consists on keeping track of the evolution of the number of connected components as the value of parameter $\epsilon$ increases; it is stored in a topological summary, formally called *persistence diagram*. Those connected components persisting for more time are the more meaningful, so they will determine the clustering structure for the given cloud point data (see Figure 1c). A persistence diagram, $T^0$, is a multiset of (*birth*, *death*) points, $(b_i, d_i)$, in $\mathbb{R}_+^2$, i.e. the first entry indicates the time (value of $\epsilon$) where a connected component is created, and the second one is its death time; so we can define this summary as

$$T^0 = \{(b_i, d_i) \in \mathbb{R}_+^2 \ : \ b_i \leq d_i, i = 1, \ldots, n\},$$

for some $m < \infty$. Black dots in Figure 1c are a graphical representation of this set for the cloud point data on Figure 1a.

This described clustering procedure, actually, corresponds to the computation of the 0-homology under the TDA terminology, and the number of connected components, or clusters, is defined as the Betty number $\beta_0$. However, persistent homology is able to model higher dimensional relationships among data points. For

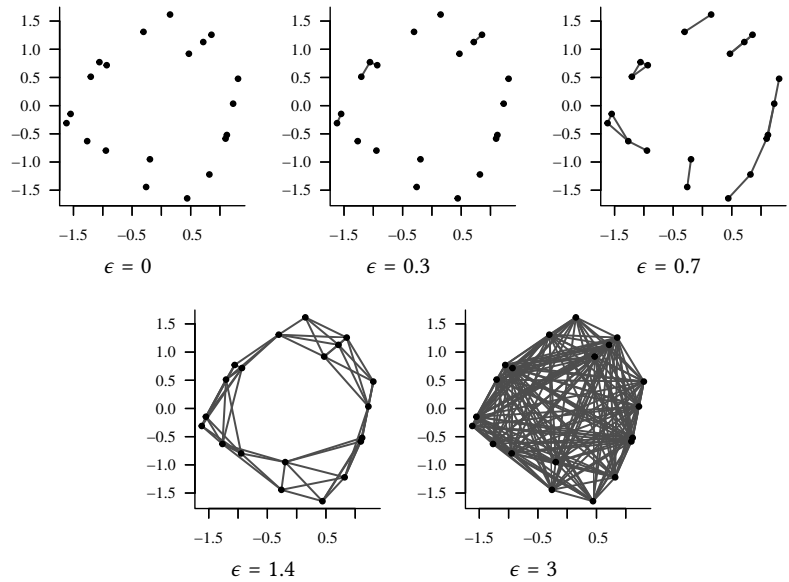

(a) Agglomerative hierarchical clustering with single linkage using different radius values $\epsilon$

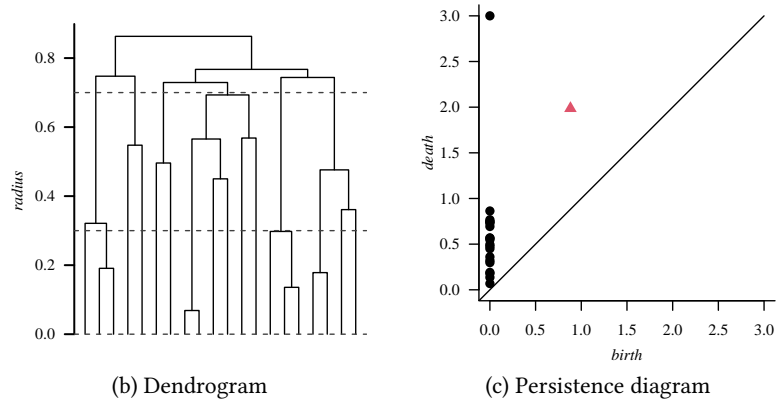

(b) Dendrogram

(c) Persistence diagram

Figure 1: Illustration of hierarchical clustering, Panels (a) and (b), and its corresponding persistence diagram, Panel (c). Horizontal dashed lines in the dendrogram correspond to some values of $\epsilon$.

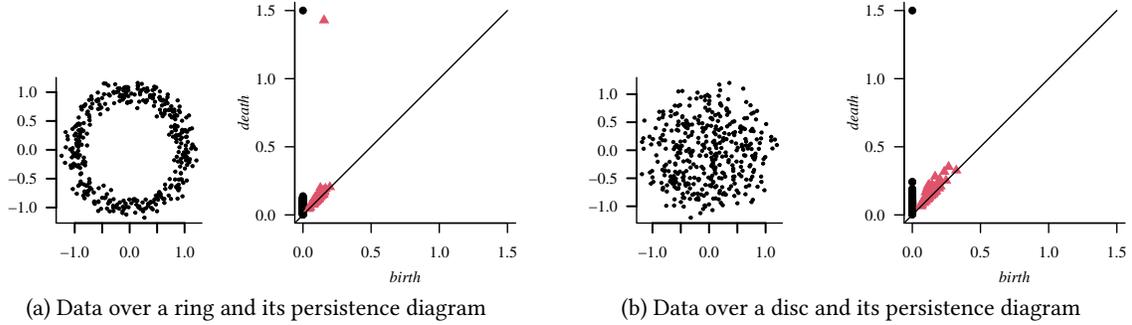

(a) Data over a ring and its persistence diagram    (b) Data over a disc and its persistence diagram

Figure 2: Illustration of 1-dimensional holes for cloud point data over different manifolds: (a) a ring, and (b) a disc.

example, Figure 2 shows two cloud point data sets, each one formed by a single cluster, but they do not have the same shape. This additional information cannot be obtained from traditional clustering procedures, but it is indeed useful to, for example, define a more adequate probabilistic model in each case. Higher levels of homology, $h$-homology for $h \geq 1$, are also captured during the computation of the persistent homology. This is one of the advantages of TDA methodologies. Betti numbers $\beta_h$, $h \geq 1$, count the number of $h$-dimensional holes. In the toy example of Figure 2, the two cloud point data sets are different by the presence of a one dimensional hole in the ring cloud point data, so we have $\beta_1 = 1$, whereas for the other one, $\beta_1 = 0$. In Figure 2, red triangles represent the 1-dimensional topological features. Using all this information together, one can have a more complete picture of the shape where data live.

Another application of TDA is related to dimensionality reduction. There are situations where each observation in a cloud point data takes values in an $s$-dimensional space, for example $\mathbb{R}^s$, but the meaningful features, or their intrinsic shape, say $\mathcal{M}$, is embedded in such a space, so $s' = \dim(\mathcal{M}) < s$. Persistence homology will only detect topological features until dimension homology $s'$, even though it is computed from the bigger space of dimension $s$.

## 3 A methodology for modeling topological features

Randomness is an important issue that has to bear in mind when performing statistical inference in persistent homology. Indeed, typically the cloud point data $y = (y_1, \ldots, y_m)$ to be subject to topological analysis is a sample of random points in a metric space. Consequently, any topological feature extracted from it has random variations that needs to be taken into consideration for meaningful inferences.

More specifically, two of the common cloud point data generating mechanisms are the following. Under the first scenario, each point $y_i$, $i = 1, \ldots, m$, is independently drawn with the same probability distribution $F$, supported on a manifold $\mathcal{M}$ embedded in $\mathbb{R}^d$. Under the second one, each point $y_i$, $i = 1, \ldots, m$, has the form $y_i = u_i + z_i$, where $u_i$ is independently drawn with the same probability distribution $F$ supported on $\mathcal{M}$, as in the first setting, and $z_i$ is an independent and identically distributed (iid) random perturbation drawn from some distribution $G$ on $\mathbb{R}^d$. For example, $G$ can be a $d$-variate Gaussian distribution with zero mean and covariance matrix $\sigma^2 I$, with $I$ the identity matrix, and the dispersion $\sigma^2 > 0$ is typically small. Another possibility is to assume that $z_i$ is randomly drawn in such a way that it is perpendicular to the tangent space of the manifold $\mathcal{M}$ at the point $u_i$. Notice that in these last two cases the cloud point data does not lie exactly over the manifold $\mathcal{M}$ but close to it.

In this framework, the main purpose of TDA is to infer topological features of the underlying manifold $\mathcal{M}$ from the given random cloud point data. In particular, as already mentioned, persistent homology keeps track of the evolution of topological features when varying a filtration parameter, in order to characterize the shape of the manifold through the evolution of its homological groups. However, this endeavor faces the difficulty that, due to the discrete nature and sampling variability of the cloud point data, topological features of short duration over the filtration emerge which are irrelevant regarding the true topology of the underlying manifold of interest $\mathcal{M}$. In [19] and [13], these features of random short lifetimes, which can be explained just by sampling variability, are termed *topological noise*, in contrast to the *topological signal* corresponding to persistent features.

Thus, discriminating between topological noise and topological signal is a crucial problem in TDA. For this, in [19] and [13], bootstrap confidence bands around the diagonal of persistent diagrams are introduced, and points (*birth*, *death*) inside the band are disregarded as due to topological noise. The so called Vietoris-Rips filtration is used to compute the persistent homology, see [16]. As described in Section 2, for each homology level $h$ the persistence diagram is represented by the multiset:

$$T^h = \{(b_i^h, d_i^h) \in \mathbb{R}_+^2 \ : \ b_i^h \leq d_i^h, i = 1, \ldots, n^h\},$$

for some $n^h < \infty$. The lifetimes of the homology classes in the persistence diagram are given by

$$l_i^h = d_i^h - b_i^h, \qquad i = 1, \ldots, n^h.$$

Randomness of the cloud point data leads to randomness of these lifetimes. Assume that $\mathcal{M}$ is a smooth manifold composed by a finite number of closed connected components, as it is common in practice. Also assume that the size $m$ of the cloud point data is large in comparison with such number of connected components, as well as with the number of holes and other Betti numbers in $\mathcal{M}$. Then, most of the lifetimes $l_i^h$, $i = 1, \ldots, n^h$, except for a few, will be the result of birth and death of homological classes due to topological noise in sampling the manifold through the cloud point data.

With all these elements explained, a general statistical modelling approach for lifetimes associated with topological noise is introduced next. This provides, as a byproduct, a statistical tool for disentangling topological noise from topological signal, without requiring intensive computations involved in bootstrapping persistent diagrams.

## 3.1 Topological signal detection through random partition modeling

Each lifetime can be identified as only one type of feature: topological noise or topological signal, and it is expected that the quantity of lifetimes being topological noise is much larger than those being topological signal. Focusing on the noise, they appear due to the sampling process, so it is possible to describe them according to some probabilistic model. In contrast, the topological signal will appear for any other reason not explained by the noise model. However, both type of features are collected together; thus, the topological signal becomes an *outlier* with respect to the topological noise. Hence, we will be able to disentangle both of them by applying some methodology designed for this purpose.

For the sake of completeness, given some arbitrary dataset, any observation which is inconsistent with the remainder is called an *outlier*. Under a probabilistic approach, it is assumed that $n - p$ observations, from a total of $n$, arise from some model, whereas the remaining $p$ observations come from a different one. In general, $n \gg p$. The literature for methods tackling the problem of outlier detection is wide and comprises computational, probabilistic and statistical approaches; see for example, [56] for a recent review, and [49, 48, 51] for some Bayesian nonparametric methodologies. In this work, a clustering approach is followed built upon Bayesian nonparametric commonly used tools, in particular, we make use of restricted random partitions as the methodological component, similarly to [23, 54].

Random partitions are probabilistic tools suitable to perform clustering [see, e.g. 33, 39], since their sampling space, known as *set partitions*, encodes every possible arrangement of any set of items into a number of nonempty groups, or clusters. As a simple example, consider the set $\{y_1, y_2, y_3\}$, then all their possible arrangements are

$$\{\{y_1, y_2, y_3\}\}, \quad \{\{y_1\}, \{y_2, y_3\}\}, \quad \{\{y_1, y_2\}, \{y_3\}\},$$
$$\{\{y_1, y_3\}, \{y_2\}\}, \quad \{\{y_1\}, \{y_2\}, \{y_3\}\},$$

where, for example, $\{\{y_1, y_2\}, \{y_3\}\}$ means that there are two clusters: one formed by observations $y_1$ and $y_2$, and the second one formed only by $y_3$. Establishing some notation, a set partition $\pi \in \mathcal{P}$ is a partition of some set of cardinality $n$ having $k$ nonempty subsets, groups or blocks, for some $1 \leq k \leq n$, where each group is denoted by $\pi_j$, $j = 1, \ldots, k$. Furthermore, simplifying the writing, partitions will be denoted by $\pi_1/\cdots/\pi_k$ instead of $\{\pi_1, \cdots, \pi_k\}$.

Outlier detection, for the context at issue, can be performed by means of a specific class of random partitions, whose support is a subset of $\mathcal{P}$. Let us assume the homology level $h$ is fixed for the rest of the explanation. Since lifetimes are all positive numbers, it will be assumed they are ordered, i.e. $l_i \leq l_{i+1}$ for $i = 1, \ldots, n - 1$, with $n$ the total number of lifetimes. This allows us to locate the potential topological signal as the largest lifetime values; the rest of them will be the topological noise. Then, it is expected that the topological signal will be grouped in a few clusters $\pi_j$, all of them with a few elements or even being singletons. By ordering lifetime values, such clusters would be the most-right of them. However, any set partition is invariant to permutations of their blocks (one of the reasons of the very well known problem of *label-switching*) and we cannot stick to this rule.

Therefore, it is necessary to restrict the sampling space of the random partition. This new space will only contain set partitions $\pi \in \mathcal{P}$ such that every block $\pi_j$ consists of consecutive items and $\max \pi_j < \min \pi_{j+1}$ for $j = 1, \ldots, k - 1$ with $k$ the number of groups in $\pi$. These conditions are known as the *no-gaps* assumption [cf. 23, 37, 54]. In the small example, the partition $\{y_1, y_3\}/\{y_2\}$ does not satisfy this assumption. Let us denote by $\mathcal{R}$ the set of all no-gaps set partitions.

Random partitions can be used in conjunction with a model-based approach, meaning that a probabilistic model $g_j$ is assigned to each group $\pi_j$. As a consequence, all observations $y_i \in \pi_j$ are distributed according to $g_j$ [iid]. Therefore, given these elements, the proposed model for outlier detection can be written hierarchically as

$$l_i | \pi, \phi \sim g(l_i | \phi_j) \mathbf{1}(l_i \in \pi_j) \quad [\text{ind}], \quad i = 1, \ldots, n, \tag{1}$$
$$\phi_j | \pi \sim \nu_0 \quad [\text{iid}],$$
$$\pi \sim \rho_0,$$

where $g$ is some probability distribution supported over $\mathbb{R}^+$ with driving finite-dimensional parameter $\phi_j$, the distribution $\nu_0$ is the prior for each parameter $\phi_j$, and $\pi$ is an $\mathcal{R}$-valued random partition with prior distribution $\rho_0$. Among the candidates for distribution $g$, the log-normal distribution of parameters $\phi_j = (\mu_j, \sigma_j^2) \in \mathbb{R} \times \mathbb{R}^+$ will be used, and its conjugate for $\nu_0$ is chosen, i.e. a normal-gamma distribution of parameters $(m, c, a, b)$. With respect to the prior distribution for the random partition, $\rho_0$, a restriction of the so-called exchangeable partition probability functions (EPPFs) is used, see [37, 54] for further details. EFFPs are a widely used class of distributions in Bayesian nonparametric methodologies. In particular, for the EPPF derived from the Dirichlet process [20], its corresponding $\mathcal{R}$-valued distribution is written as

$$\mathbf{P}(\pi = \pi_1/\cdots/\pi_k) = \binom{n}{n_1 \cdots n_k} \frac{\theta^k}{k!(\theta)_{n\uparrow}} \prod_{j=1}^{k} \Gamma(n_j), \tag{2}$$

for any set partition $\pi = \pi_1/\ldots/\pi_k \in \mathcal{R}$, and where $n_j = \#\pi_j$ is the cardinality of block $\pi_j$ for $j = 1\ldots, k$, $(x)_{r\uparrow} = x(x+1)\cdots(x+r-1)$ is the Pochhammer symbol or rising factorial, and $\theta > 0$ is the total mass parameter for the Dirichlet process.

In order to compute point estimates for this model, we resort to numerical procedures, in particular, Markov chain Monte Carlo (MCMC) techniques. Given the data, a sample of the posterior distribution

$$p(\pi|l_1, \ldots, l_n) \propto \rho_0(\pi) L(l_1, \ldots, l_n | \pi), \tag{3}$$

is obtained. In Appendix A, the derivation of the complete MCMC sampling scheme is presented. For an alternative estimation procedure based on neural networks, see [24].

## 3.2 Estimation of Betti numbers

Working under the $\mathcal{R}$ approach allows us to locate the potential topological signal as the rightmost groups of a partition $\pi$, whereas the rest of them will contain the topological noise. Topological noise might contain some information regarding the underlying geometry of the sampling space, whereas the topological signal is of interest since it determines the value for the Betti numbers $\beta_h$, for each homology level $h \geq 0$. Given our probabilistic framework, we are only able to provide some point estimate $\hat{\beta}_h$, $h \geq 0$.

Perhaps, the most natural way to estimate the Betti numbers is using the posterior modal partition, say $\tilde{\pi}$. Assuming $\tilde{\pi}$ has $k$ groups, block $\tilde{\pi}_k$ will contain the maximum lifetime value, $l_n$. If there is some topological signal, by definition, it is encoded in the largest lifetimes, in particular $l_n$. Thus, the number of elements in block $\tilde{\pi}_k$ serves an estimate for the Betti number $\beta$, i.e. $\hat{\beta} = \#\tilde{\pi}_k$.

Model (1) resembles mixture models [see, e.g. 39, 36, for more details]. One of the advantages of mixture models is their capability for fitting complex distributions, which is achieved by combining several mixing components, even though the resulting distribution exhibits a single mode. This is noteworthy to say since it might be expected that the topological noise behaves according to some probability distribution, even though it is not bounded to be of the form of a single kernel function $g(\cdot|\phi_j)$. As a result, the posterior modal partition $\tilde{\pi}$ can be conformed by more than two clusters, which ideally would be one for each topological feature (noise and signal). According to the definition of outlier, we can only expect that the size of $\tilde{\pi}_k$ is very small when compared with the sample size $n$. Therefore, a possible drawback of using only the last block $\tilde{\pi}_k$ to estimate Betti numbers is that the topological signal may be spread over a few blocks $\{\pi_j : j \geq k^*\}$, for some $k^*$ very close to $k$.

In order to overcome this possible underestimation and determine from which block of $\tilde{\pi}$ outliers start appearing, the degree of overlap between each pair of consecutive mixing components, $w_j g(\cdot|\phi_j)$ and $w_{j+1} g(\cdot|\phi_{j+1})$, for $j = 1, \ldots, k - 1$, can be measured. It is expected that the overlapping for those components modeling the topological noise is bigger than the one with the component modeling the topological signal. In particular, assuming blocks $\{\pi_j : j \geq k^*\}$ contain the topological signal, the degree of overlap between $w_{k^*-1} g(\cdot|\phi_{k^*-1})$ and $w_{k^*} g(\cdot|\phi_{k^*})$ should be small. We use the measure in [58]

$$\Delta(f_1, f_2) = \int \min(f_1(x), f_2(x)) dx, \tag{4}$$

to quantify the degree of overlap.

On the other hand, another estimation procedure can be used. From the posterior distribution for $\pi$, Equation (3), the marginal probability, $S_i$, that each lifetime $l_i$ starts a group can be computed, which is given by

$$S_i = \sum_{\pi \in \mathcal{R}} \mathbf{P}(\min \pi_j = l_i \text{ for some } j = 1, \ldots, k | \cdots),$$

for $i = 2, \ldots, n$. A similar point estimate is used in [34, 37] under a context of change point detection. Rightmost lifetimes values having a high probability can be considered as the starting point of the topological signal. Therefore, a second point estimate for Betti numbers, denoted by $\check{\beta}$, can be defined as

$$\check{\beta} = n - \max\{i = 2, \ldots, n : S_i \geq p_0\} + 1,$$

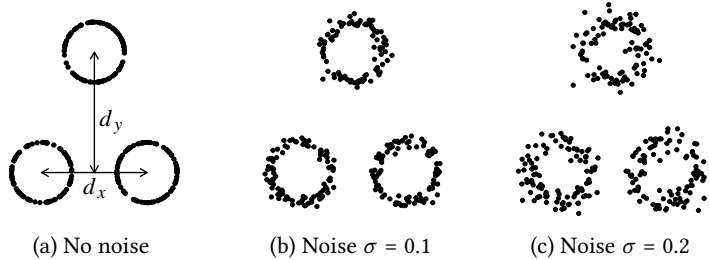

(a) No noise   (b) Noise $\sigma = 0.1$   (c) Noise $\sigma = 0.2$

Figure 3: Examples for the different noise levels using the same manifold. Distances, displayed in Panel (a), $d_x$ and $d_y$ will range from 1 to 5 units.

for some threshold value $p_0 \in [0, 1]$. First lifetime, $l_1$, is excluded since it always starts a group.

For the 0-homology level, it is important to highlight that the value obtained for $\hat{\beta}_0$ should be corrected by adding one; similarly, for $\check{\beta}_0$. The computation of persistence diagrams requires a maximum radius $\epsilon$, which is somehow arbitrarily fixed, and there will always be a lifetime having such a value. Then, it is necessary to remove it before any further processing, but it does count one connected component.

## 4 Simulation study

The proposed methodology is tested under some simulated scenarios, using manifolds having different shape and dimension. We start considering the 0th homology level using synthetic cloud point data uniformly distributed over the circle, varying the quantity of circles, their location, and the sample size. Additionally, a small noise is introduced in each dataset (see Figure 3) with the purpose of better understand the robustness of the model.

First examples are taken from a manifold conformed by $r$ circles, $r = 1, 2, 3$; a cloud point data of size $n = 600$ is drawn for each case. In addition, two levels of noise are included, as depicted in Figure 3, consisting on a Gaussian perturbation of each point over the circle with standard deviation $\sigma = 0.1, 0.2$. For the cases $r = 2, 3$, the circles are separated from their centers by 5 units, according to Figure 3a. An MCMC run was executed for each cloud data point, taking a sample of size 5 000 after discarding a first batch of 10 000. Hyperparameter settings are as follows: $(0, 0.5, 1.1, 0.1)$ for the prior distribution $v_0$, and $(1.1, 0.1)$ for the total mass parameter, $\theta$, prior. Table 1 and Figure 4 present the posterior estimates for these datasets. In all cases, the value of $\beta_0$ is correctly estimated using both approaches, i.e. $\hat{\beta}_0$ and $\check{\beta}_0$ are the same. Posterior modal partition, reported in terms of block sizes, detects some groups of noise lifetimes, but the last one is much smaller than the rest for the two- and three-circles examples, so it contains the topological signal. The degree of overlapping between the last two mixing components helps to assess this fact. On the other hand, marginal probabilities $S_i$ besides to provide the same information, their graphical form (Figure 4) shows some other relevant features in the data. The supplemental material contains a more exhaustive simulation study.

A second example consists on a manifold made by spheres whose centers are points of a spherical Fibonacci lattice. Over each sphere, a random sample of 50 points is drawn; the manifold consists on 30 spheres, see Figure 5. The purpose of this example is testing the estimation of higher dimensional holes, namely $\beta_1$ and $\beta_2$. The MCMC specification is the same as before. Posterior estimates are presented in Table 2 and Figure 6. Regarding $\beta_0$, one connected component is lost in both estimates; this might be due to the data sampling process. In fact, a closer inspection of the persistence diagram shows that there are only 28 persistent features. For the one-dimensional holes, $\beta_1$, no persistent features are detected. Finally, the study of two-dimensional holes shows an interesting fact. The highest marginal probability is $S_{47}$ indicating one void, the one due to the Fibonacci lattice, but the second higher probability is $S_{17}$ meaning that the void for each individual small sphere is also detected. These results for $H_1$ and $H_2$ are consistent with the topological features of a sphere, i.e. $\beta_1 = 0$ and $\beta_2 = 1$, even if $S_{17}$ is considered as the estimate for $\beta_2$ since the manifold would be conformed by 30 small spheres plus the bigger one. On the other hand, it is worth saying that for each homology level, a different number of topological features are detected: $n^0 = 1\,199$ plus the one removed as explained, $n^1 = 330$, and $n^2 = 47$; except for $n^0$ which is always the number of observations, the rest is random.

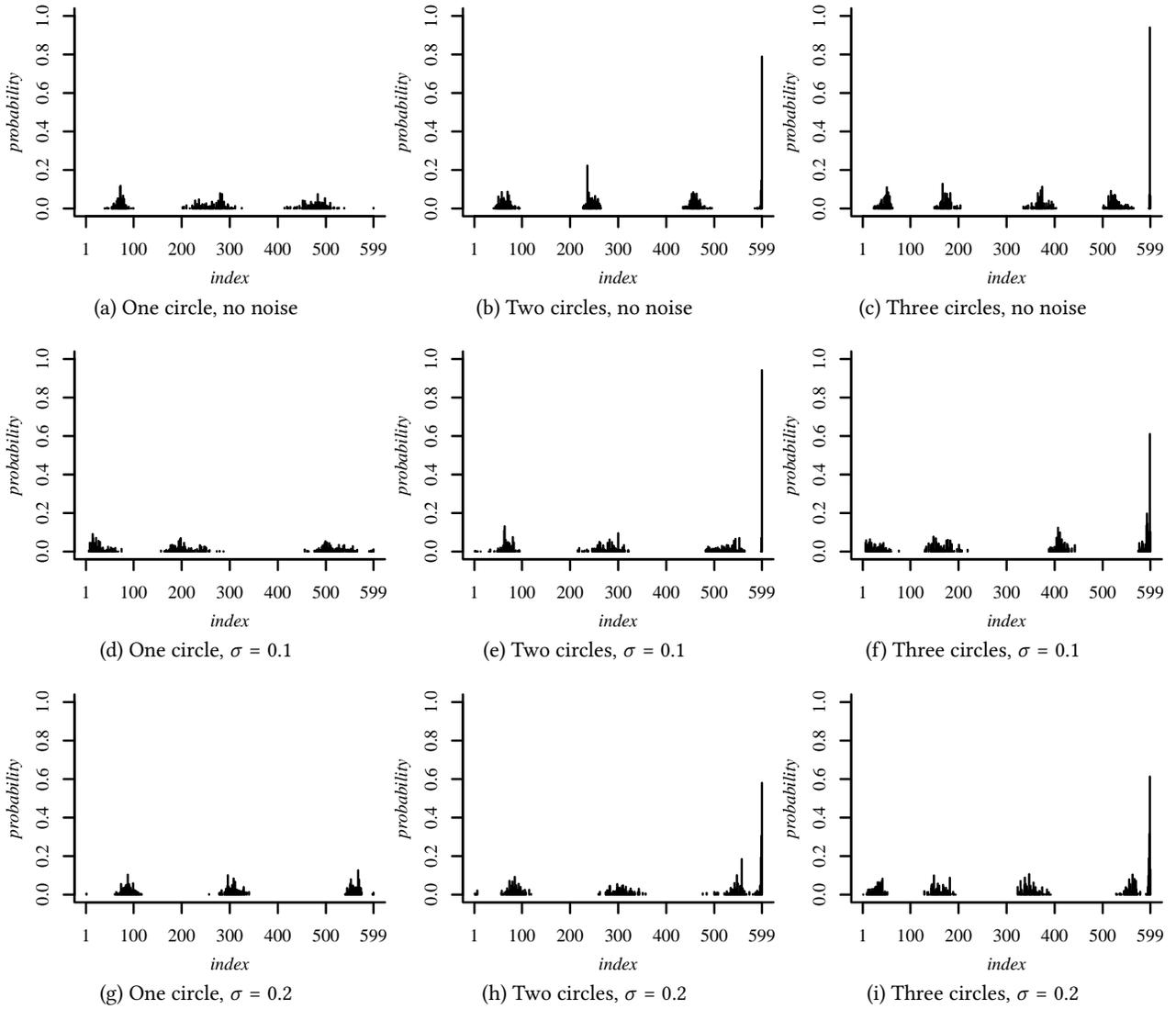

Figure 4: Marginal probability, $S_i$, for each lifetime $l_i$ to start a group for the different toy examples.

Table 1: Posterior estimates for the firsts toy examples, consisting on $r$ circles, with Gaussian noise $\sigma$. Modal partition $\tilde{\pi}$ is presented in terms of block sizes $(n_1, \ldots, n_k)$ together with its probability. Given $\tilde{\pi}$, the degree of overlap is presented in the fifth column. Next, the indices $i$ and their marginal probabilities $S_i$ for each lifetime greater than $p_0 = 0.3$ are shown. Last columns contain the estimated Betti numbers.

| $r$ | $\sigma$ | $(n_1, \ldots, n_k)$ | prob. | $\{\Delta(\hat{f}_j, \hat{f}_{j+1}) : j = 1, \ldots, k-1\}$ | $\{i \geq 2 : S_i \geq 0.30\}$ | probs. | $\hat{\beta}_0$ | $\check{\beta}_0$ |
|---|---|---|---|---|---|---|---|---|
| 1 | — | (72, 207, 203, 117) | 0.019 | {0.084, 0.214, 0.154} | {} | — | 1 | 1 |
|   | 0.1 | (14, 167, 311, 107) | 0.017 | {0.015, 0.192, 0.139} | {} | — | 1 | 1 |
|   | 0.2 | (87, 208, 256, 48) | 0.018 | {0.107, 0.238, 0.055} | {} | — | 1 | 1 |
| 2 | — | (69, 168, 211, 150, 1) | 0.019 | {0.084, 0.179, 0.177, 0.001} | {599} | {0.789} | 2 | 2 |
|   | 0.1 | (73, 226, 241, 58, 1) | 0.017 | {0.087, 0.243, 0.073, 0.001} | {599} | {0.942} | 2 | 2 |
|   | 0.2 | (89, 220, 247, 42, 1) | 0.017 | {0.106, 0.227, 0.048, 0.001} | {598, 599} | {0.305, 0.580} | 2 | 2 |
| 3 | — | (39, 127, 204, 154, 73, 2) | 0.022 | {0.048, 0.140, 0.187, 0.098, 0.002} | {598} | {0.940} | 3 | 3 |
|   | 0.1 | (15, 137, 254, 174, 17, 2) | 0.001 | {0.016, 0.156, 0.209, 0.019, 0.003} | {598} | {0.611} | 3 | 3 |
|   | 0.2 | (40, 130, 179, 216, 32, 2) | 0.010 | {0.049, 0.161, 0.205, 0.036, 0.003} | {597, 598} | {0.315, 0.613} | 3 | 3 |

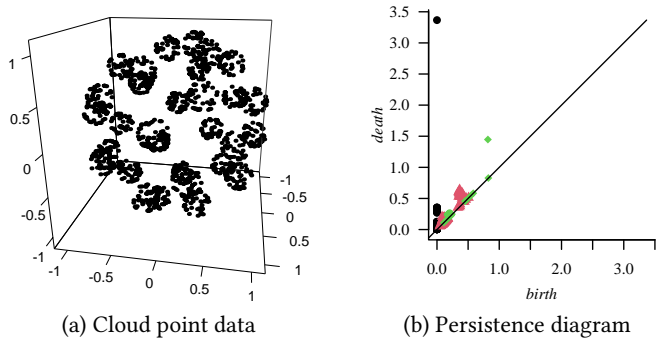

(a) Cloud point data   (b) Persistence diagram

Figure 5: Cloud point data and persistence diagram over a spherical-Fibonacci manifold.

## 5 Concluding remarks

Topological data analysis is an emerging field of applied mathematics providing useful topological and geometrical information about the sample space. In particular, we have described persistent homology, one of the most common methodologies in TDA, where the main purpose is to discover topological features in such a space. The most relevant features are summarized in the Betti numbers $\beta_h$, $h \geq 0$, quantifying the number of $h$ dimensional holes.

Under a statistical viewpoint, determining the values for the Betti numbers is an inference problem. However, it has not been straightforward providing point estimates. The persistence diagram, the topological summary of persistent homology, takes values in a very complex space. Additionally, the observed cloud point data contains an inherent randomness, which is translated to the persistence diagram, so not all recorded features are relevant. Therefore, in any persistence diagram, topological noise and topological signal are mixed up.

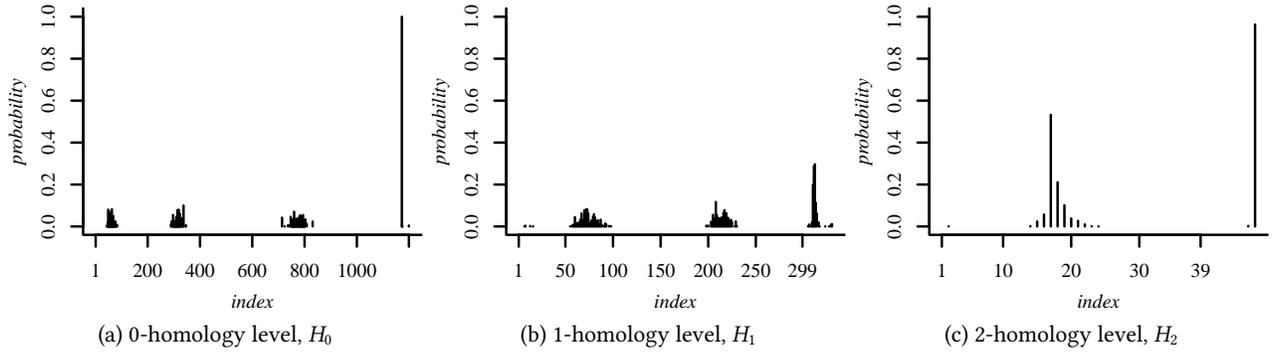

Figure 6: Marginal probability, $S_i$, for each lifetime $l_i$ to start a group for the spherical-Fibonacci manifold example, for the homology levels $H_0$, $H_1$, and $H_2$.

Table 2: Posterior estimates for the spherical-Fibonacci manifold example, for the homology levels $H_0$, $H_1$, and $H_2$. Modal partition $\tilde{\pi}$ is presented in terms of block sizes $(n_1, \ldots, n_k)$ together with its probability. Given $\tilde{\pi}$, the degree of overlap is presented in the fifth column. Next, the indices $i$ and their marginal probabilities $S_i$ for each lifetime greater than $p_0 = 0.6$ are shown. Last columns contain the estimated Betti numbers.

| $h$ | $(n_1, \ldots, n_k)$ | prob. | $\{\Delta(\hat{f}_j, \hat{f}_{j+1}) : j = 1, \ldots, k-1\}$ | $\{i \geq 2 : S_i \geq 0.6\}$ | probs. | $\hat{\beta}_h$ | $\check{\beta}_h$ |
|---|---|---|---|---|---|---|---|
| 0 | (76, 258, 468, 369, 28) | 0.020 | {0.046, 0.152, 0.215, 0.007} | {1172} | {1} | 29 | 29 |
| 1 | (72, 143, 96, 19) | 0.021 | {0.150, 0.203, 0.041} | {} | {} | — | — |
| 2 | (16, 30, 1) | 0.521 | {0.209, 0.020} | {47} | {0.963} | 1 | 1 |

While this work aims to close the gap between TDA and Statistics practitioners, its main contribution is to provide a statistical study of persistence diagrams by means of lifetime's topological features. This approach eases the disentanglement of the topological features and allows to identify and quantify the topological signal. Following a full Bayesian framework, the topological signal identification is treated as an outlier detection problem. The presented clustering model, based on random partitions, agglomerates the most persistent lifetimes as outliers, and their number is associated with the corresponding Betti number. Further, the methodology is tested by an extensive simulation study.

Moreover, the posterior estimates seem to preserve the geometric information gathered by persistent homology as can be seen in the marginal probabilities. A deep study along these results, and their applications, are part of the ongoing work.

## A  MCMC sampling scheme

Assume there is a random sample $y = (y_1, \ldots, y_n)$ where $y_i \in \mathbb{R}^+$ and $y_i \leq y_{i+1}$ for $i = 1, \ldots, n-1$. Let $\pi$ be an $\mathcal{R}$-valued random partition having $k$ groups for some $1 \leq k \leq n$. According to Model (1), the joint posterior distribution for $(\phi, \pi)$ is given by

$$p(\phi, \pi | y) \propto \ell(y|\phi, \pi) p(\phi, \pi) = \ell(y|\phi, \pi) p(\phi|\pi) \rho_0(\pi),$$

with $\ell$ the likelihood function

$$\ell(y|\phi, \pi) = \prod_{j=1}^{k} \prod_{i \in \pi_j} g(y_i | \phi_j),$$

and $p(\phi|\pi)$ the prior joint distribution for kernel parameters, written as $p(\phi|\pi) = \prod_{j=1}^{k} v_0(\phi_j)$. Since our interest is on the marginal posterior distribution of $\pi$, kernel parameters are integrated out in what is known as the marginal likelihood, i.e.

$$L(y|\pi) = \prod_{j=1}^{k} \int_{\Phi} \prod_{i \in \pi_j} g(y_i | \phi) v_o(\mathrm{d}\phi).$$

The log-normal distribution of parameters $\phi = (\mu, \tau)$ is used as the kernel function $g$ to model lifetimes $l_i$. It is known that if $X$ is a random variable log-normal distributed, then $Y = \ln X$ is normally distributed. Thus, let $y_i = \ln l_i$ for $i = 1, \ldots, n$. Additionally, with respect to the prior distribution for $\phi$, a conjugate normal-gamma of parameters $(m, c, a, b)$ is chosen, i.e.

$$\mu | \tau \sim \mathrm{N}(\mu | m, c/\tau), \qquad \tau \sim \mathrm{Ga}(\tau | a, b),$$

for $m \in \mathbb{R}$ and $a, b, c > 0$. Therefore, the marginal likelihood takes the form

$$L(y|\pi) = \prod_{j=1}^{k} \frac{b^a \Gamma\left(\frac{n_j}{2} + a\right)}{(2\pi)^{n_j/2} (n_j c + 1)^{1/2} \Gamma(a)} \left\{ \frac{S_{\pi_j}}{2} + \frac{n_j (\bar{y}_{\pi_j} - m)^2}{2(n_j c + 1)} + b \right\}^{-\left(\frac{n_j}{2} + a\right)}, \qquad (5)$$

with $n_j = \#\pi_j$ the cardinality of block $\pi_j$, $j = 1 \ldots, k$,

$$\bar{y}_{\pi_j} = \frac{1}{n_j} \sum_{i \in \pi_j} y_i \quad \text{and} \quad S_{\pi_j} = \sum_{i \in \pi_j} (y_i - \bar{y}_{\pi_j})^2.$$

The posterior distribution (3) is, thus, obtained from the marginal likelihood in (5) and the prior for $\pi$ given in Equation (2).

In order to draw samples from the posterior, we make use of MCMC techniques. Due to the no-gaps assumption for the possible groupings, it is possible to explore the space $\mathcal{R}$ through an MCMC sampler belonging to the category *split-merge*. Since any group $\pi_j$, $j = 1, \ldots, k$ in partition $\pi$ consists of consecutive observations, the sampler can propose to perform a split, for example, of group $\pi_r = \{i_1, i_2, \ldots, i_{n_r}\}$ into two, $\pi'_r$ and $\pi'_{r+1}$, such that $\pi'_r = \{i_1, \ldots, i_s\}$ and $\pi'_r = \{i_{s+1}, \ldots, i_r\}$ for some $s = 2, \ldots, n_r - 1$. The opposite move consists, then, in selecting two adjacent groups, say $\pi_r$ and $\pi_{r+1}$, and merge them into a single one, that is $\pi'_r = \pi_r \cup \pi_{r+1}$. Each move is chosen randomly, and then, the proposed operation is accepted through a Metropolis-Hasting step. This algorithm is proposed by [23], and the full details can be also found in [37].

After drawing the value for the partition $\pi$, other parameters can be updated in the sampling scheme. In this case, we include the total mass parameter $\theta$, since it can highly influence the number of groups. Following [18], if a gamma prior of parameters $(\alpha, \beta)$ is assigned, its conditional posterior distribution is given by

$$p(\theta|\eta, k) = q\text{Ga}(\theta|\alpha + k, \beta - \log \eta) + (1 - q)\text{Ga}(\theta|\alpha + k - 1, \beta - \log \eta), \qquad (6)$$

for $\eta \sim \text{Be}(\theta + 1, n)$, $k$ is the current number of groups, and

$$\frac{q}{1-q} = \frac{\alpha + k - 1}{n(\beta - \log \eta)}.$$

## A.1 Kernel parameter sampling

Once a partition point estimate is obtained, the posterior mode $\tilde{\pi}$ for example, observations in the same block $\tilde{\pi}_j$ follow the same model $g(\cdot|\phi_j)$ for $j = 1, \ldots, \#\tilde{\pi}$. In order to compute the degree of overlap (4), an estimate for each kernel parameter $\phi_j$ is required. For the log-normal case and the given prior, the full conditional distributions for $\phi = (\mu, \tau)$ are conjugate, that is a normal-gamma of parameters $(m', c', a', b')$ such that:

$$m' = \frac{nc\bar{y} + m}{nc + 1}, \qquad\qquad a' = \frac{n}{2} + a,$$

$$c' = \frac{c}{nc + 1}, \qquad\qquad b' = \frac{S}{2} + \frac{n(\bar{y} - m)^2}{2(nc + 1)} + b,$$

with $\bar{y}$ the sample mean, and $S = \sum_{i=1}^{n}(y_i - \bar{y})$.

# Extended simulation study for the estimation of topological features

Asael Fabian Martínez

Universidad Autónoma Metropolitana, Unidad Iztapalapa, Mexico City, Mexico

fabian@xanum.uam.mx

An extensive simulation study was done using different synthetic cloud point data in order to understand the performance of the proposed methodology. Based on the scenarios described in Section 4, the sample size will be set to $n = 300$, $600$, and $900$, and the separation of circles ranges from 1 to 5 for the cases $r = 2, 3$. For each cloud point data, $s = 100$ replications were simulated, and each estimator for $\beta_0$ was computed. Tables 1–7 present the error in the estimation, that is

$$\text{er}(\cdot) = \frac{1}{s} \sum_{j=1}^{s} \mathbf{1}(\cdot \neq \beta_0),$$

for $\beta_0$ the true value for each point cloud data, which is the number of circles $r$.

From the results, we can see that the method is able to correctly estimate the number of connected components, $\beta_0$, in most of the cases for $r = 1$. As expected, the noise makes harder the estimation. For the cases $r = 2, 3$, the separation of the circles is also crucial for a correct estimation.

Table 1: Estimation error for each estimator, $\hat{\beta}_0$ and $\check{\beta}_0$ for $r = 1$ circle and for the different sample sizes $n$. Here $\beta_0 = 1$. Variable $\sigma$ indicates the noise level in terms of the standard deviation, with '—' indicating no noise was added.

| | \multicolumn{6}{c}{$n$} | | | | |
|---|---|---|---|---|---|---|
| | 300 | | 600 | | 900 | |
| $\sigma$ | $\text{er}(\hat{\beta}_0)$ | $\text{er}(\check{\beta}_0)$ | $\text{er}(\hat{\beta}_0)$ | $\text{er}(\check{\beta}_0)$ | $\text{er}(\hat{\beta}_0)$ | $\text{er}(\check{\beta}_0)$ |
| — | 0.00 | 0.00 | 0.00 | 0.00 | 0.00 | 0.04 |
| 0.1 | 0.00 | 0.02 | 0.08 | 0.41 | 0.02 | 0.50 |
| 0.2 | 0.16 | 0.42 | 0.07 | 0.40 | 0.08 | 0.63 |

Table 2: Estimation error for each estimator, $\hat{\beta}_0$ and $\check{\beta}_0$ for $r = 2$ circles, for the different sample sizes $n$, and separation $d_x$ between centers' circles. Here $\beta_0 = 2$. Variable $\sigma$ indicates the noise level in terms of the standard deviation, with '—' indicating no noise was added.

|  |  | \multicolumn{6}{c}{$n$} |  |  |  |  |
| --- | --- | --- | --- | --- | --- | --- | --- |
|  |  | \multicolumn{2}{c}{300} | \multicolumn{2}{c}{600} | \multicolumn{2}{c}{900} |
| $d_x$ | $\sigma$ | er($\hat{\beta}_0$) | er($\check{\beta}_0$) | er($\hat{\beta}_0$) | er($\check{\beta}_0$) | er($\hat{\beta}_0$) | er($\check{\beta}_0$) |
| 1 | — | 1.00 | 1.00 | 1.00 | 0.99 | 1.00 | 1.00 |
|  | 0.1 | 1.00 | 0.98 | 1.00 | 0.98 | 0.97 | 0.95 |
|  | 0.2 | 0.95 | 0.91 | 0.95 | 0.93 | 0.95 | 0.96 |
| 2 | — | 1.00 | 0.99 | 1.00 | 1.00 | 1.00 | 1.00 |
|  | 0.1 | 1.00 | 1.00 | 0.97 | 0.98 | 0.98 | 0.97 |
|  | 0.2 | 0.96 | 0.86 | 0.94 | 0.93 | 0.95 | 0.98 |
| 3 | — | 0.00 | 0.01 | 0.00 | 0.00 | 0.00 | 0.01 |
|  | 0.1 | 0.01 | 0.12 | 0.03 | 0.21 | 0.19 | 0.34 |
|  | 0.2 | 0.85 | 0.84 | 0.94 | 0.97 | 0.97 | 0.98 |
| 4 | — | 0.00 | 0.00 | 0.00 | 0.01 | 0.00 | 0.01 |
|  | 0.1 | 0.00 | 0.01 | 0.00 | 0.07 | 0.00 | 0.06 |
|  | 0.2 | 0.00 | 0.18 | 0.00 | 0.14 | 0.00 | 0.10 |
| 5 | — | 0.00 | 0.06 | 0.00 | 0.04 | 0.00 | 0.03 |
|  | 0.1 | 0.00 | 0.03 | 0.00 | 0.13 | 0.01 | 0.11 |
|  | 0.2 | 0.00 | 0.17 | 0.00 | 0.08 | 0.00 | 0.09 |

Table 3: Estimation error for each estimator, $\hat{\beta}_0$ and $\check{\beta}_0$ for $r = 3$ circles, for the different sample sizes $n$, and separation $d_x$ between centers' circles. Here $\beta_0 = 3$, and $d_y = 1$. Variable $\sigma$ indicates the noise level in terms of the standard deviation, with '—' indicating no noise was added.

| | | \multicolumn{6}{c}{$n$} | | | | | |
|---|---|---|---|---|---|---|---|
| | | \multicolumn{2}{c}{300} | \multicolumn{2}{c}{600} | \multicolumn{2}{c}{900} |
| $d_x$ | $\sigma$ | er($\hat{\beta}_0$) | er($\check{\beta}_0$) | er($\hat{\beta}_0$) | er($\check{\beta}_0$) | er($\hat{\beta}_0$) | er($\check{\beta}_0$) |
| 1 | — | 1.00 | 1.00 | 1.00 | 1.00 | 1.00 | 1.00 |
|   | 0.1 | 1.00 | 1.00 | 1.00 | 0.98 | 1.00 | 0.99 |
|   | 0.2 | 1.00 | 0.98 | 0.98 | 0.96 | 0.98 | 0.99 |
| 2 | — | 1.00 | 1.00 | 1.00 | 1.00 | 1.00 | 1.00 |
|   | 0.1 | 1.00 | 0.98 | 1.00 | 0.99 | 1.00 | 0.99 |
|   | 0.2 | 0.98 | 0.99 | 0.98 | 0.91 | 1.00 | 0.99 |
| 3 | — | 1.00 | 1.00 | 1.00 | 1.00 | 1.00 | 1.00 |
|   | 0.1 | 1.00 | 0.99 | 1.00 | 0.98 | 1.00 | 0.99 |
|   | 0.2 | 1.00 | 1.00 | 1.00 | 0.97 | 1.00 | 0.98 |
| 4 | — | 1.00 | 1.00 | 0.98 | 0.89 | 0.46 | 0.66 |
|   | 0.1 | 1.00 | 1.00 | 1.00 | 0.98 | 1.00 | 0.98 |
|   | 0.2 | 1.00 | 0.97 | 0.97 | 0.97 | 1.00 | 1.00 |
| 5 | — | 0.03 | 0.07 | 0.00 | 0.01 | 0.00 | 0.02 |
|   | 0.1 | 0.82 | 0.81 | 0.65 | 0.70 | 0.80 | 0.85 |
|   | 0.2 | 1.00 | 0.96 | 0.99 | 0.99 | 1.00 | 1.00 |

Table 4: Estimation error for each estimator, $\hat{\beta}_0$ and $\check{\beta}_0$ for $r = 3$ circles, for the different sample sizes $n$, and separation $d_x$ between centers' circles. Here $\beta_0 = 3$, and $d_y = 2$. Variable $\sigma$ indicates the noise level in terms of the standard deviation, with '—' indicating no noise was added.

| | | \multicolumn{6}{c}{$n$} | | | | | |
|---|---|---|---|---|---|---|---|
| | | \multicolumn{2}{c}{300} | \multicolumn{2}{c}{600} | \multicolumn{2}{c}{900} |
| $d_x$ | $\sigma$ | er($\hat{\beta}_0$) | er($\check{\beta}_0$) | er($\hat{\beta}_0$) | er($\check{\beta}_0$) | er($\hat{\beta}_0$) | er($\check{\beta}_0$) |
| 1 | — | 1.00 | 1.00 | 1.00 | 1.00 | 1.00 | 1.00 |
| | 0.1 | 1.00 | 1.00 | 1.00 | 0.98 | 1.00 | 0.98 |
| | 0.2 | 0.99 | 0.97 | 0.99 | 0.99 | 1.00 | 0.99 |
| 2 | — | 1.00 | 1.00 | 1.00 | 1.00 | 0.99 | 0.97 |
| | 0.1 | 1.00 | 1.00 | 1.00 | 0.99 | 1.00 | 0.99 |
| | 0.2 | 1.00 | 0.98 | 1.00 | 0.98 | 0.99 | 0.99 |
| 3 | — | 0.33 | 0.50 | 0.04 | 0.12 | 0.00 | 0.11 |
| | 0.1 | 1.00 | 0.98 | 0.99 | 0.98 | 0.99 | 0.97 |
| | 0.2 | 1.00 | 0.98 | 0.98 | 0.97 | 1.00 | 1.00 |
| 4 | — | 0.04 | 0.03 | 0.01 | 0.00 | 0.00 | 0.02 |
| | 0.1 | 0.40 | 0.45 | 0.30 | 0.40 | 0.50 | 0.65 |
| | 0.2 | 1.00 | 0.95 | 0.99 | 0.96 | 1.00 | 1.00 |
| 5 | — | 0.04 | 0.03 | 0.01 | 0.00 | 0.00 | 0.01 |
| | 0.1 | 0.02 | 0.02 | 0.07 | 0.10 | 0.08 | 0.13 |
| | 0.2 | 0.71 | 0.75 | 0.91 | 0.91 | 0.94 | 0.94 |

Table 5: Estimation error for each estimator, $\hat{\beta}_0$ and $\check{\beta}_0$ for $r = 3$ circles, for the different sample sizes $n$, and separation $d_x$ between centers' circles. Here $\beta_0 = 3$, and $d_y = 3$. Variable $\sigma$ indicates the noise level in terms of the standard deviation, with '—' indicating no noise was added.

| | | \multicolumn{6}{c}{$n$} | | | | | |
|---|---|---|---|---|---|---|---|
| | | \multicolumn{2}{c}{300} | \multicolumn{2}{c}{600} | \multicolumn{2}{c}{900} |
| $d_x$ | $\sigma$ | er($\hat{\beta}_0$) | er($\check{\beta}_0$) | er($\hat{\beta}_0$) | er($\check{\beta}_0$) | er($\hat{\beta}_0$) | er($\check{\beta}_0$) |
| 1 | — | 1.00 | 1.00 | 1.00 | 1.00 | 1.00 | 1.00 |
| | 0.1 | 0.99 | 0.98 | 0.99 | 0.96 | 0.97 | 0.94 |
| | 0.2 | 0.99 | 0.90 | 1.00 | 0.98 | 1.00 | 0.98 |
| 2 | — | 1.00 | 1.00 | 1.00 | 0.99 | 1.00 | 1.00 |
| | 0.1 | 1.00 | 0.96 | 0.99 | 0.97 | 0.99 | 0.92 |
| | 0.2 | 0.97 | 0.96 | 0.99 | 0.97 | 0.99 | 0.99 |
| 3 | — | 0.28 | 0.07 | 0.05 | 0.02 | 0.02 | 0.04 |
| | 0.1 | 0.51 | 0.11 | 0.35 | 0.13 | 0.49 | 0.26 |
| | 0.2 | 0.87 | 0.85 | 0.97 | 0.93 | 0.98 | 0.98 |
| 4 | — | 0.00 | 0.02 | 0.00 | 0.00 | 0.00 | 0.02 |
| | 0.1 | 0.01 | 0.01 | 0.00 | 0.01 | 0.01 | 0.01 |
| | 0.2 | 0.31 | 0.19 | 0.36 | 0.29 | 0.39 | 0.36 |
| 5 | — | 0.00 | 0.02 | 0.00 | 0.02 | 0.00 | 0.01 |
| | 0.1 | 0.00 | 0.02 | 0.00 | 0.01 | 0.00 | 0.04 |
| | 0.2 | 0.12 | 0.10 | 0.15 | 0.08 | 0.11 | 0.09 |

Table 6: Estimation error for each estimator, $\hat{\beta}_0$ and $\check{\beta}_0$ for $r = 3$ circles, for the different sample sizes $n$, and separation $d_x$ between centers' circles. Here $\beta_0 = 3$, and $d_y = 4$. Variable $\sigma$ indicates the noise level in terms of the standard deviation, with '—' indicating no noise was added.

| | | \multicolumn{6}{c}{$n$} | | | | | |
|---|---|---|---|---|---|---|---|
| | | \multicolumn{2}{c}{300} | \multicolumn{2}{c}{600} | \multicolumn{2}{c}{900} |
| $d_x$ | $\sigma$ | er($\hat{\beta}_0$) | er($\check{\beta}_0$) | er($\hat{\beta}_0$) | er($\check{\beta}_0$) | er($\hat{\beta}_0$) | er($\check{\beta}_0$) |
| 1 | — | 1.00 | 0.97 | 1.00 | 0.95 | 1.00 | 0.99 |
| | 0.1 | 1.00 | 0.99 | 1.00 | 0.98 | 1.00 | 0.95 |
| | 0.2 | 1.00 | 0.97 | 0.99 | 0.96 | 0.99 | 0.96 |
| 2 | — | 1.00 | 0.98 | 1.00 | 0.99 | 1.00 | 1.00 |
| | 0.1 | 1.00 | 0.99 | 1.00 | 0.97 | 1.00 | 0.96 |
| | 0.2 | 1.00 | 0.93 | 1.00 | 0.97 | 1.00 | 0.97 |
| 3 | — | 1.00 | 0.19 | 0.98 | 0.18 | 0.96 | 0.08 |
| | 0.1 | 1.00 | 0.23 | 0.98 | 0.21 | 0.94 | 0.28 |
| | 0.2 | 1.00 | 0.88 | 1.00 | 0.95 | 1.00 | 0.96 |
| 4 | — | 0.00 | 0.02 | 0.00 | 0.01 | 0.00 | 0.02 |
| | 0.1 | 0.00 | 0.04 | 0.00 | 0.03 | 0.00 | 0.04 |
| | 0.2 | 0.10 | 0.11 | 0.15 | 0.10 | 0.13 | 0.18 |
| 5 | — | 0.00 | 0.04 | 0.00 | 0.02 | 0.00 | 0.03 |
| | 0.1 | 0.00 | 0.05 | 0.00 | 0.08 | 0.00 | 0.03 |
| | 0.2 | 0.00 | 0.10 | 0.00 | 0.12 | 0.00 | 0.10 |

Table 7: Estimation error for each estimator, $\hat{\beta}_0$ and $\check{\beta}_0$ for $r = 3$ circles, for the different sample sizes $n$, and separation $d_x$ between centers' circles. Here $\beta_0 = 3$, and $d_y = 5$. Variable $\sigma$ indicates the noise level in terms of the standard deviation, with '—' indicating no noise was added.

| | | \multicolumn{6}{c}{$n$} | | | | | |
|---|---|---|---|---|---|---|---|
| | | \multicolumn{2}{c}{300} | \multicolumn{2}{c}{600} | \multicolumn{2}{c}{900} |
| $d_x$ | $\sigma$ | er($\hat{\beta}_0$) | er($\check{\beta}_0$) | er($\hat{\beta}_0$) | er($\check{\beta}_0$) | er($\hat{\beta}_0$) | er($\check{\beta}_0$) |
| 1 | — | 1.00 | 0.99 | 1.00 | 0.97 | 1.00 | 0.96 |
| | 0.1 | 1.00 | 0.93 | 0.99 | 0.97 | 0.99 | 0.91 |
| | 0.2 | 1.00 | 0.88 | 1.00 | 0.95 | 0.99 | 0.93 |
| 2 | — | 1.00 | 0.96 | 1.00 | 0.96 | 1.00 | 0.96 |
| | 0.1 | 1.00 | 0.99 | 1.00 | 0.96 | 0.99 | 0.94 |
| | 0.2 | 1.00 | 0.93 | 1.00 | 0.94 | 1.00 | 0.93 |
| 3 | — | 1.00 | 0.35 | 1.00 | 0.30 | 1.00 | 0.21 |
| | 0.1 | 1.00 | 0.29 | 1.00 | 0.34 | 0.92 | 0.46 |
| | 0.2 | 1.00 | 0.84 | 1.00 | 0.97 | 1.00 | 0.93 |
| 4 | — | 0.00 | 0.04 | 0.00 | 0.02 | 0.00 | 0.06 |
| | 0.1 | 0.00 | 0.06 | 0.00 | 0.08 | 0.00 | 0.11 |
| | 0.2 | 0.32 | 0.23 | 0.43 | 0.22 | 0.52 | 0.23 |
| 5 | — | 0.00 | 0.08 | 0.00 | 0.01 | 0.00 | 0.04 |
| | 0.1 | 0.00 | 0.04 | 0.00 | 0.11 | 0.01 | 0.15 |
| | 0.2 | 0.00 | 0.09 | 0.00 | 0.10 | 0.00 | 0.13 |